# Observation of anomalous spin-torque generated by a ferromagnet

A. Bose*[1], D. D. Lam[2], S. Bhuktare[1], S. Dutta[1], H. Singh[1], S. Miwa[2], A. A. Tulapurkar[1]

[1]*Department of Electrical Engineering, Indian Institute of Technology Bombay, Powai, Mumbai – 400 076, India*
[2]*Graduate School of Engineering Science, Osaka University, Toyonaka, Osaka 560-8531, Japan*
*<u>arnabbose@ee.iitb.ac.in</u>

Over the years central research of spintronics has focused on generating spin-current to manipulate nano-magnets by spin torque. So far electrically[1–9] and thermally driven spin-torques[10–12] have been experimentally demonstrated. These torques can be attributed to either Slonczewski's spin-transfer torque (STT)[13] or field-like torque (FLT)[3,8,14]. STT arises when ferromagnet absorbs spin current generated by many ways like spin-Hall effect (SHE)[15,16], spin-pumping[17], spin-Nernst effect[18], spin-(dependent) Seebeck effect[19–21] etc. Field-like torque is generally observed in asymmetric magnetic tunnel junctions (MTJ) with current perpendicular to the plane (CPP) geometry[3,14,22] and ferromagnet/heavy metal bilayer where Rashba[8] or Dressulhous[7] spin orbit interaction is present. Control of magnetization dynamics is not only interesting from physics perspective but also useful in technological applications[23,24]. We have experimentally observed a new form of spin torque which is completely different from conventional STT and FLT. This unconventional spin torque is exerted by a fixed magnet on a free magnet in spin valve structure with current in-plane (CIP) geometry. The observed spin torque originates from an out of plane effective magnetic field with symmetry of ($M \times J$), where $M$ denotes the magnetization direction of fixed FM and J denotes current density. This torque could be potentially useful for switching out-of-plane magnets in high density MRAM.

Spin-orbit torque[4,7,8,25] has evolved in a promising way to manipulate spins since last few years. Heavy metals like Pt[26], anti-ferromagnets[27,28], two dimensional materials[29,30] and semiconductor systems[7] have been recently studied as candidates for generating spin torques. However ferromagnetic metal (FM) itself has its own spin-orbit coupling which is responsible for various effects like: anisotropic magneto resistance (AMR), planar Hall effect (PHE) and anomalous Hall effect (AHE). AHE is analogous to SHE in heavy metal which can induce spin current in neighbouring metal and cause spin orbit torque (Fig. 1b). Previous studies show that spin Hall angle of FM[31,32] is quite comparable to Pt. Hence FM can be considered as good candidate for SOT[33-34]. To study spin orbit torque by FM we need FM(free)/Cu/FM(fixed) heterostructure where fixed layer will be source of spin current which will exert torque on another FM separated by Cu spacer. Based on this principle we carried out spin-torque ferromagnetic resonance (ST-FMR)[3,7,26] measurement of current in-plane giant-magnetoresistive (GMR) stack consisting of Ta(5 nm)/Ru(5 nm)/IrMn(7 nm)/CoFe(2 nm)/Cu(5 nm)/CoFe(2 nm)/Cu(5 nm). However we surprisingly observe the existence of a new kind of torque which is completely different from standard spin-orbit torque by FM (owing to its AHE) which we initially expected. This unconventional spin-torque depends on the mutual orientation of fixed layer magnetization direction ($M$) and direction of in-plane current flow ($J$) and manifests itself as an effective magnetic field perpendicular to M and J ($H^{\perp} \propto (M \times J)$). This is also markedly different from the spin torques observed in current perpendicular to plane (CPP) devices, where the spin torque depends only on the angle between the directions of free and fixed layer magnetizations (Fig. 1a).

We fabricated GMR stack as shown in Fig. 1c. Magnetization of bottom ferromagnetic CoFe was pinned by annealing the sample at 300°C for two hours in external in-plane magnetic field of 0.6 T. Top CoFe (2 nm) is free layer which is separated from bottom fixed magnetic layer by 5 nm thick Cu spacer. Top CoFe (free layer) is protected by 5 nm of Cu cap. The stack was pattered to rectangular shape (375 μm × 25 μm) by optical lithography and argon ion milling. Radio frequency (RF) current is passed in plane of the GMR stack and voltage is measured through an inductor of a bias-tee (Fig. 1e). Frequency is swept in presence of constant external field. For FM/HM structure generally magnetic field is swept to measure ST-FMR[7,26,28] but frequency sweep[3,5,6,14] is favourable in case of GMR and MTJ structure since fixed layer can also move at higher field which can lead to erroneous result. Resistance of the sample depends on the mutual orientation of magnetization between free and fixed layer (Fig. 1d) which originates from current in-plane GMR effect[21]. Clear GMR hysteresis is observed (Fig. 1d) which dominates over other effects like AMR etc. On application of RF current if spin torque (STT, FLT, new kind of torque, Oersted field-torque) is applied on free layer it undergoes precession which results in oscillation of resistance due to the GMR effect. Homodyne mixture of RF current and oscillatory resistance can result in large DC voltage at resonance of free layer[3]. Since we have dominant GMR (Fig. 1d) effect we expect maximum signal when angle between free and fixed layer is 90 degree and signal should vanish when free and pinned layers are parallel or anti-parallel.

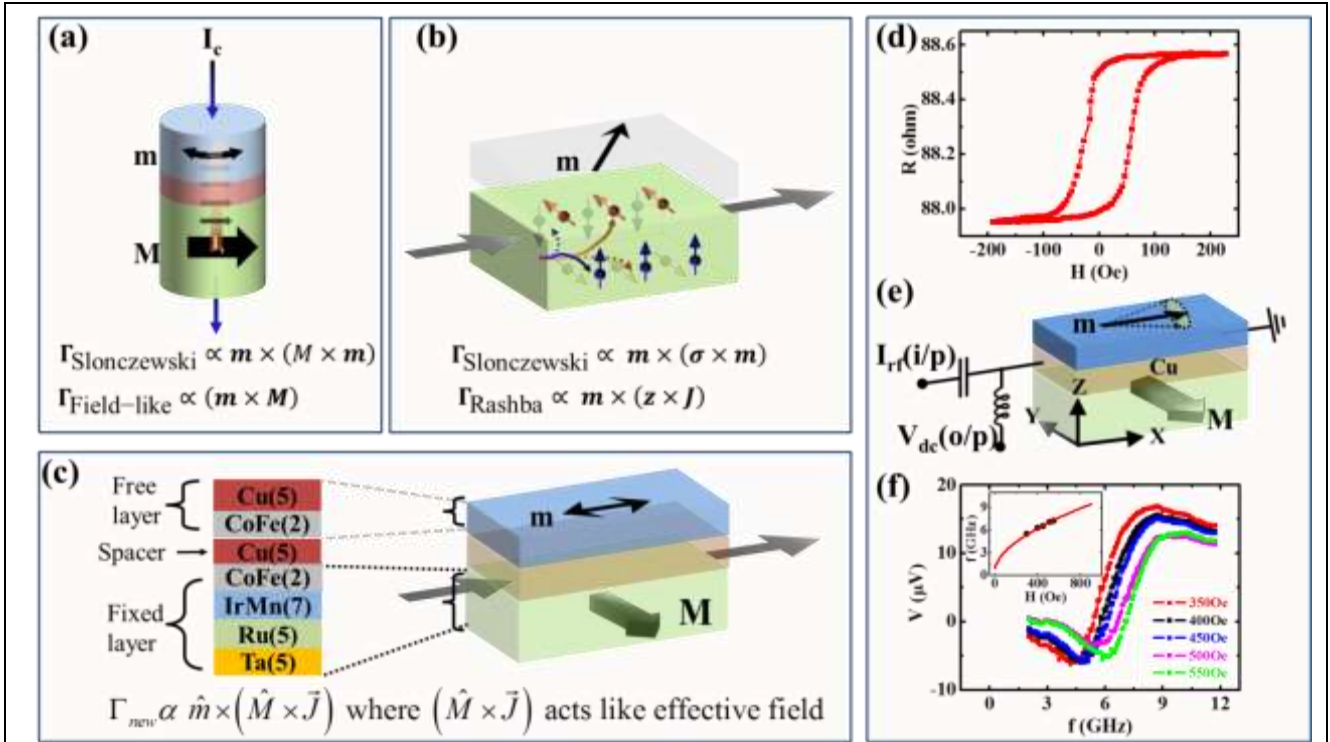

Figure 1| (**a**) Conventional GMR/MTJ stack where current flows perpendicular to plane resulting Slonczewski-like spin-transfer torque (STT) and field-like torque (FLT). (**b**) Typical bi-layer structure consisting of heavy metal (HM) and ferromagnet (FM). Spin-torque is produced on FM due to bulk charge to spin conversion by SHE (STT) or interfacial Rashba spin-orbit coupling (Rashba-like FLT). (**c**) Schematic of a GMR stack where in plane current can produce out-of-plane effective field ($H^\perp$) driven torque on free layer where $H^\perp$ field is determined by the mutual orientation of applied current and magnetization of fixed layer $H^\perp \alpha$ (**M×J**). (**d**) Magnetoresistance of GMR stack when magnetic field is applied parallel to pinned layer. (**e**) Schematic diagram of ST-FMR experimental set up. (**f**) ST-FMR signal of CIP-GMR stack while angle between free and fixed layer is 90 degree and applied power 18 dBm. Inset of (**f**) shows the Kittel's fit of resonance frequency as a function of magnetic field applied along *X*-axis.

Figure 1.e-f show the experimental set up and results of dc voltage produced by ST-FMR of current in-plane (CIP) GMR respectively. Clear dc voltage consisting of symmetric Lorentzian $\left(V_S = C_1 \frac{\Delta^2}{4(f-f_0)^2 + \Delta^2}\right)$ and anti-symmetric Lorentzian $\left(V_A = C_2 \frac{4(f-f_0)\Delta}{4(f-f_0)^2 + \Delta^2}\right)$ is observed for different values of external magnetic field (Fig. 1f). Resonance frequency, $f_0$ as a function of applied magnetic field was obtained by fitting the data to a combination of symmetric Lorentzian and anti-symmetric Lorentzian. Resonance frequency can be fitted well to Kittel's formula: $f_0 = \frac{\gamma}{2\pi}\sqrt{(H_P + H_{Ext})((H_P + H_{Ext} + H_\perp))}$ (inset of Fig. 1f) where $H_P$ is in-plane anisotropy field (~50 Oe), $H_\perp$ is out of plane anisotropy field ($1.35 \times 10^4$ Oe), $H_{ext}$ is external applied field and $\gamma$ is gyromagnetic ratio ($2.05 \times 10^5$ A/m-sec). Anti-symmetric component of dc voltage arises from Oersted field as more current flows below the free layer. The symmetric Lorentzian

component can arise from in-plane polarized spin-current absorbed by free layer (Fig. 1a,b) or in-plane current driven out-of-plane magnetic field (Fig. 1c). Studying angular dependence of ST-FMR we can ascribe that observed symmetric component in our experiment originates from in-plane current driven out-of-plane magnetic field ($H^\perp \propto (M \times J)$). We use the co-ordinate system as shown in Fig. 1e. The sample is in $X$-$Y$ plane and RF current flows along $X$ axis (Fig. 1e). The equilibrium magnetization direction of free layer ($m$) and pinned layer ($M$) are in $X$-$Y$ plane and make angles of $\theta_m$ and $\theta_M$ with respect to $X$ axis (direction of RF current flow).

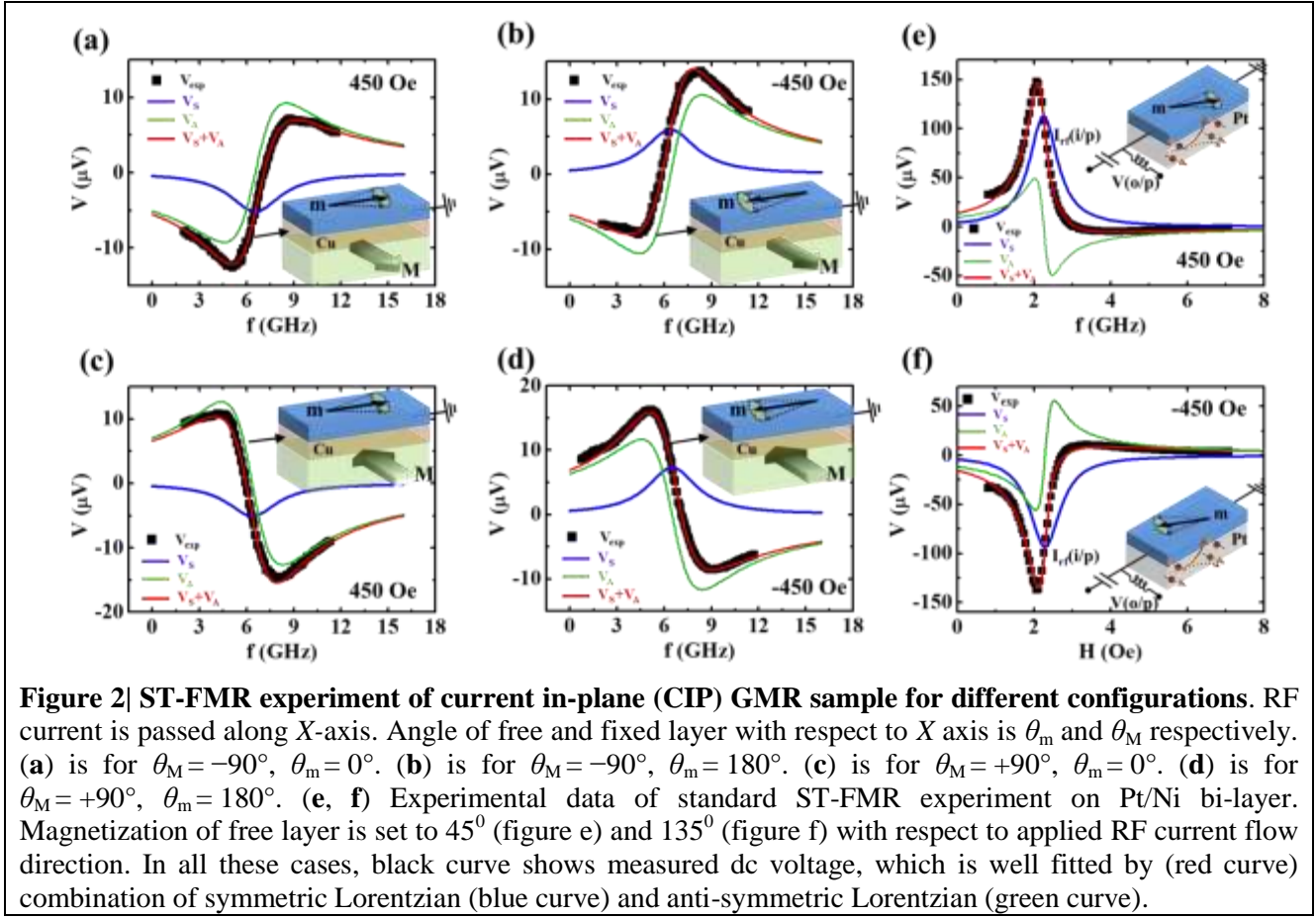

**Figure 2| ST-FMR experiment of current in-plane (CIP) GMR sample for different configurations.** RF current is passed along $X$-axis. Angle of free and fixed layer with respect to $X$ axis is $\theta_m$ and $\theta_M$ respectively. (**a**) is for $\theta_M = -90°$, $\theta_m = 0°$. (**b**) is for $\theta_M = -90°$, $\theta_m = 180°$. (**c**) is for $\theta_M = +90°$, $\theta_m = 0°$. (**d**) is for $\theta_M = +90°$, $\theta_m = 180°$. (**e**, **f**) Experimental data of standard ST-FMR experiment on Pt/Ni bi-layer. Magnetization of free layer is set to $45^0$ (figure e) and $135^0$ (figure f) with respect to applied RF current flow direction. In all these cases, black curve shows measured dc voltage, which is well fitted by (red curve) combination of symmetric Lorentzian (blue curve) and anti-symmetric Lorentzian (green curve).

To understand the behaviour of spin-torque generated by fixed magnet, we studied detailed angular dependence as shown in figure 2 and 4. Figure 2a and 2b show ST-FMR spectrum for external field $H$=450 Oe along $X$ axis and along $-X$ axis respectively while direction of fixed layer is along $-Y$ axis ($\theta_M = -90°$). Background voltage has been subtracted from the experimental dc voltage data as shown by the black curve in all figures. (DC voltage with both free and pinned layer magnetizations along $X$-axis is taken as background (See supplementary information)). Fitted red curve is sum of symmetric Lorentzian (blue curve) and anti-symmetric Lorentzian (green curve). We can see from Fig 2a and 2b that, in both these cases anti-symmetric component ($V_A$) remains the same which is consistent with ST-FMR measurement with GMR effect detection, but surprisingly symmetric component ($V_S$) inverts its sign. This cannot be explained if we assume that pinned layer generates spin-current due to its spin-Hall effect, which is then absorbed by the free layer. This point can be appreciated by comparing results of standard ST-FMR experiment[26] on Ni/Pt sample shown in figure 2.e and 2.f. The magnetic field in this case was applied at 45° (fig 2.e) and 135° (fig 2.f) respectively. In ST-FMR experiment if detection method is based on AMR then both $V_A$ (which arises from the Oersted magnetic field of current) and $V_S$ (which arises from spin-current generated via spin-Hall effect) should invert the sign simultaneously on reversal of external field (Fig 2.e-f)[26] whereas if detection method is GMR then there should not be any sign change of $V_A$ and $V_S$ (see supplementary information). In the experiment of in-plane GMR sample we observe $V_A$ does not change its sign which is quite expected but $V_S$ changes the sign on reversal of external magnetic field. Next we measured ST-FMR signal of the GMR stack for applied field along $\pm X$ axis with pinned layer magnetization along $+Y$ axis (Fig. 2c and 2d). When external magnetic field direction is fixed but pinned layer magnetization is inverted we observe that sign of $V_A$ is changed which is quite expected (see supplementary information) but $V_S$ remains same (compare Fig. 2a with Fig. 2c and Fig. 2b with Fig. 2d). This also cannot be explained by conventional STT and FLT terms. In fact, the data shown in Fig. 2c (or Fig. 2d) can be obtained from Fig. 2a (or Fig. 2b) by rotating the co-ordinate system 180 degree (i.e.: $V_{dc}$(panel c) = $-V_{dc}$(panel b) and $V_{dc}$(panel d) = $-V_{dc}$(panel a)).

This anomalous result observed in GMR sample can be explained if we assume an out-of-plane effective magnetic field of the form $\boldsymbol{H}^\perp \propto (\boldsymbol{M} \times \boldsymbol{J})$ is created which exerts torque on the free layer (see supplementary information). An obvious conclusion from above relation is that if the current flows along the pinned layer magnetization, there should not be any out-of-plane field and hence symmetric component in the dc voltage should vanish. We verified this experimentally, and results are shown in Fig. 3a. It can be seen from figure 3.a that the symmetric component is very small in comparison to the anti-symmetric component. We can also have another interesting configuration where the dc voltage contains only the symmetric Lorentzian term but no dispersion. Such a configuration is shown in Fig. 3b, where the pinned layer makes 45° angle with $X$-axis and free layer magnetization is along $Y$-axis. In this case the Oersted magnetic field is parallel to the equilibrium magnetization direction of free layer. Hence it does not excite FMR, whereas the anomalous magnetic field ($\boldsymbol{H}^\perp$) along $Z$-axis is non-zero and excites FMR. From figure 3b we can see that the anti-symmetric component ($V_A$) is very small in comparison to the symmetric component ($V_S$). A third special case is where both $V_A$ and $V_S$ should vanish in principle when fixed layer magnetization is parallel to current flow ($X$-axis) and free layer magnetization points perpendicular to this ($Y$-axis). In this case though the GMR detection is active, as both the Oersted field and anomalous field fail to excite FMR, no dc voltage is expected. Such case is shown in figure 3.c where dc voltage signal is fairly small compared to figure 2. A small non zero signal is observed due to slight tilting of fixed layer when external field is perpendicular to it (see supplementary information). These results strongly support the existence of an anomalous in-plane current driven out-of-plane magnetic field in GMR sample. In a control experiment where fixed layer and free layer are parallel ($\theta_M = 45°$, $\theta_m = 45°$ degree) we do not observe any ST-FMR signal (Fig. 3d). This shows that our detection method is based on resistance variation of the sample by in-plane GMR effect. We have performed additional experiments to further substantiate the existence of anomalous spin-torque. Figure 4 shows the angular dependence of ST-FMR data where magnetization of pinned layer is along $-Y$ direction and external magnetic field of 450 Oe is applied along various values angles ($\theta_H$) with respect to $X$-axis. As the magnetic field value is much larger than the coercive field (~50 Oe), the free layer magnetization direction is almost same as external field direction. It evident that signal is maximum when angle between free and pinned layer is near 90 degree and signal vanishes when it is 0 degree. It is quite consistent with the theory that symmetric component has $\cos\theta_m$ dependence (Fig 4c) and anti-symmetric component has $\cos^2\theta_m$ (Fig 4d) dependence if we incorporate such out of plane effective field ($\boldsymbol{H}^\perp \propto (\boldsymbol{M} \times \boldsymbol{J})$). However a small deviation is expected if we consider slight rotation of fixed layer in response to applied external magnetic field (See detailed explanation in the supplementary information).

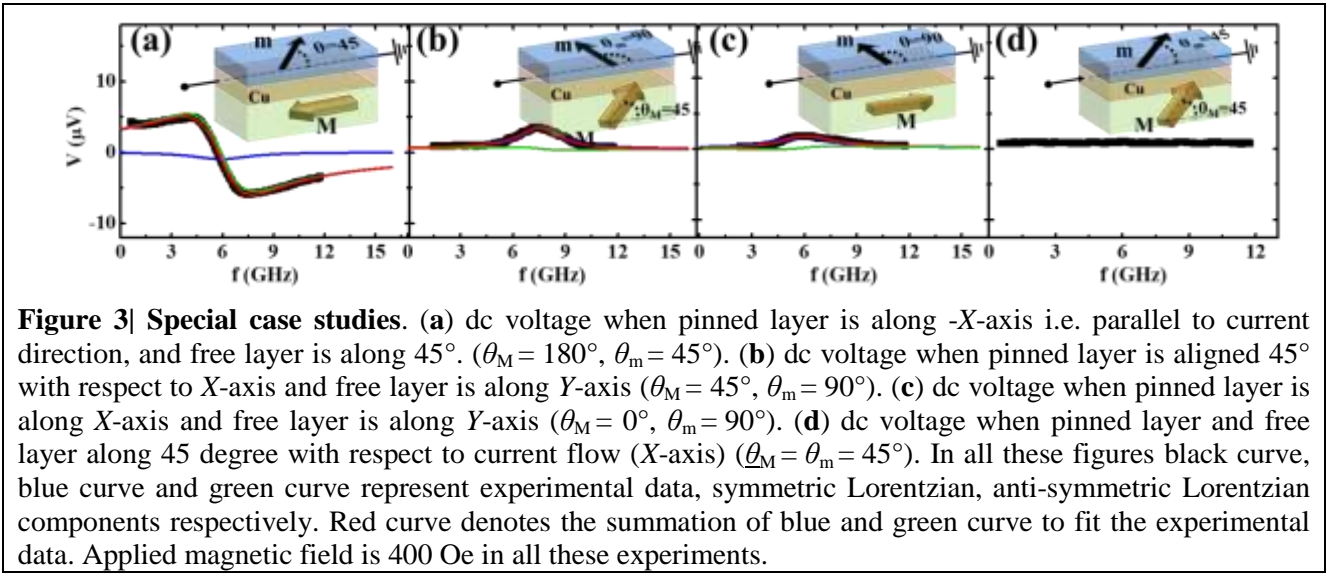

**Figure 3| Special case studies**. (**a**) dc voltage when pinned layer is along -X-axis i.e. parallel to current direction, and free layer is along 45°. ($\theta_M = 180°$, $\theta_m = 45°$). (**b**) dc voltage when pinned layer is aligned 45° with respect to X-axis and free layer is along Y-axis ($\theta_M = 45°$, $\theta_m = 90°$). (**c**) dc voltage when pinned layer is along X-axis and free layer is along Y-axis ($\theta_M = 0°$, $\theta_m = 90°$). (**d**) dc voltage when pinned layer and free layer along 45 degree with respect to current flow (X-axis) ($\theta_M = \theta_m = 45°$). In all these figures black curve, blue curve and green curve represent experimental data, symmetric Lorentzian, anti-symmetric Lorentzian components respectively. Red curve denotes the summation of blue and green curve to fit the experimental data. Applied magnetic field is 400 Oe in all these experiments.

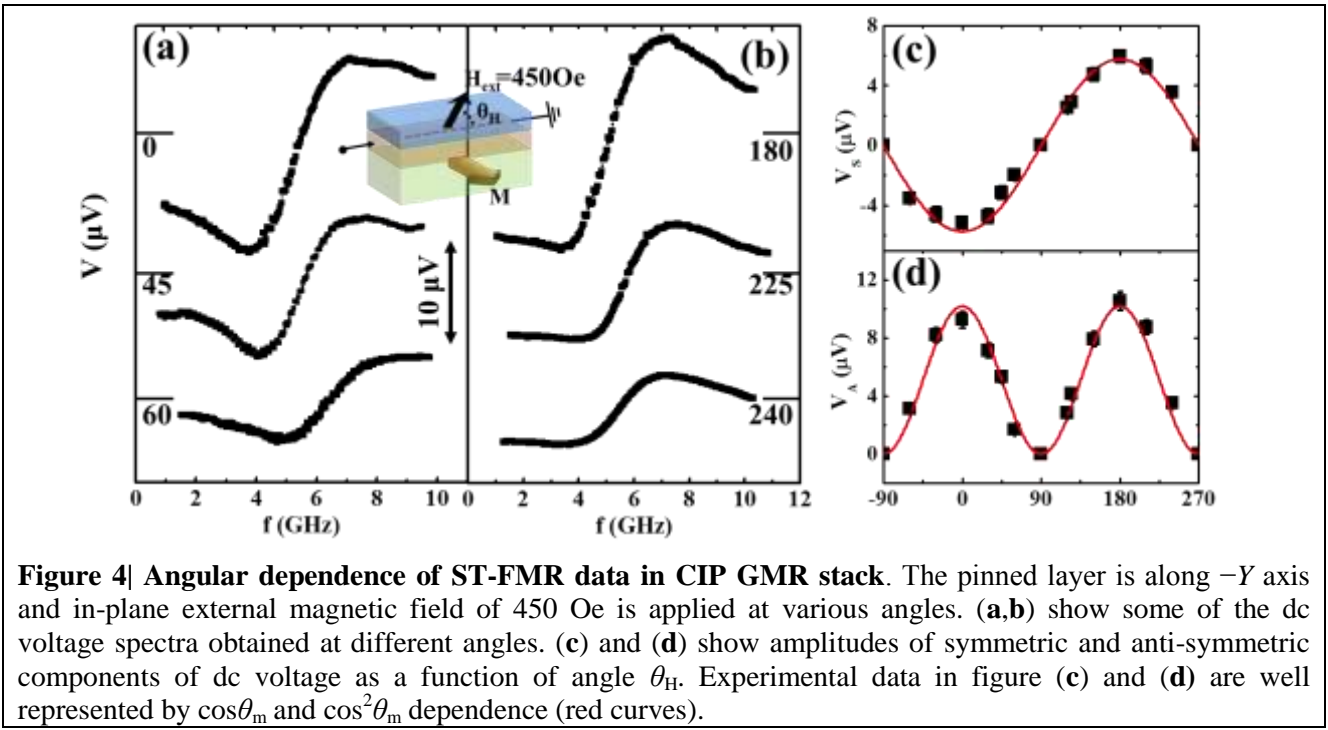

**Figure 4| Angular dependence of ST-FMR data in CIP GMR stack**. The pinned layer is along $-Y$ axis and in-plane external magnetic field of 450 Oe is applied at various angles. (**a,b**) show some of the dc voltage spectra obtained at different angles. (**c**) and (**d**) show amplitudes of symmetric and anti-symmetric components of dc voltage as a function of angle $\theta_H$. Experimental data in figure (**c**) and (**d**) are well represented by $\cos\theta_m$ and $\cos^2\theta_m$ dependence (red curves).

Recently it is proposed that in GMR kind of structure, fixed layer can produce torque on free layer owing to anomalous Hall effect and AMR of FM[33, 34]. (The AMR effect is not expected to produce spin torque with in-plane fixed layer.) However in GMR kind of structure, fixed layer and free layer are not decoupled from each other, as far as in-plane current flow is concerned. Interfacial scattering and reflection of spins take place at the interface through the Cu spacer due to zigzag motion of carriers which causes the in-plane GMR effect. So we cannot think a simple picture of GMR stack where current flows in parallel channels through the fixed layer, free layer and Cu spacer unlike HM/FM bi-layers. If the fixed FM behaves similar to other heavy metal (Pt, Ta, W etc.) we would expect injection of spin-current from fixed FM to free FM through the spin-transport via Cu spacer. In that situation we would not see sign reversal of symmetric component while reversing the external field as observed here. It is possible that current in-plane GMR effect (interfacial spin scattering and zigzag motion of electrons between fixed and free layer through the Cu spacer) in combination with spin-orbit coupling could produce such current induced out-of-plane effective field which acts on free layer as spin-torque. We have estimated that approximately 125 Oe effective out-of-plane magnetic field is created when average $10^{12}$ A/m$^2$ current density flows in FM (top = 2 nm)/Cu (5 nm)/FM (bottom = 2nm) heterostructure (details in the Supplementary Information). A detailed microscopic theory of this torque is lacking here which remains an open question.

In summary we have reported unprecedented observation of current induced out-of-plane field generated spin-torque in current-in-plane (CIP) GMR structure. This kind of spin torque can be controlled by manipulating the fixed layer magnetization direction. Such a torque can be highly useful to manipulate perpendicular magnetic bits.

**Author Contributions**

GMR film was deposited by DL with supervision from YS and SM. The lithographic device fabrication and measurements were carried out by AB with help from SB, SD and HS. AB analysed the data and wrote the manuscript with help from AT. AT supervised the project. All authors contributed to this work and commented on this paper.

**Acknowledgement:** We are thankful to the Centre of Excellence in Nanoelectronics (CEN) at the IIT-Bombay Nanofabrication facility (IITBNF) and Department of Electronics and Information Technology (DeitY), Government of India for its support. We also thank to JSPS KAKENHI (JP26103002).

Competing Interests: The authors declare that they have no competing interests.

# Supplementary information:

**S1. Derivation of the expression for dc voltage:**

The reference frame used is shown below.

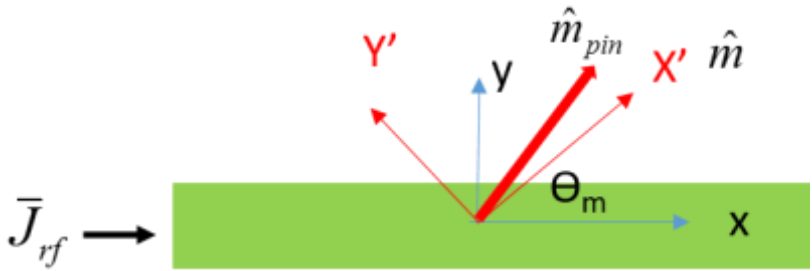

Figure S1| Estimation of ST-FMR voltage

The green rectangle denotes the GMR stack cut into rectangular shape. The rf current flows along *X*-axis. The equilibrium free layer and pinned layer magnetization directions ($\hat{m}$ and $\hat{m}_{pin}$) are assumed to be in *X-Y* plane. X'-axis is taken to be along $\hat{m}$. When current is passed along x direction, it creates Oersted magnetic field along y-axis, $\bar{h} = h_{Oe}\,\hat{y}$. The anomalous magnetic field is given by, $\bar{h}_{new} = \beta(\hat{m}_{pin} \times \bar{J}) = -\beta J\, m_{pin,y}\,\hat{z}$, where $\beta$ is a constant and $m_{pin,Y}$ denotes *Y* component of $\hat{m}_{pin}$. Assuming small oscillation of $\hat{m}$, we get the following equations:

$$\delta m_{y'} = \chi_{11} h_{y'} + \chi_{12} h_z = \chi_{11} h_{Oe}\cos\theta_m - \beta J\, m_{pin,y}\,\chi_{12} \quad ---(1)$$
$$\Rightarrow \delta m_x = -\sin\theta_m \delta m_{y'} = (-\sin\theta_m \cos\theta_m\,\chi_{11} h_{Oe} - \sin\theta_m\,\beta\, m_{pin,y}\,\chi_{12} J) \,---(2)$$
$$\text{and}\quad \delta m_y = \cos\theta_m \delta m_{y'} = (\cos^2\theta_m\,\chi_{11} h_{Oe} - \cos\theta_m\,\beta\, m_{pin,y}\,\chi_{12} J) ---(3)$$

Where $\chi$ denotes the podar susceptibility tensor.

The sample resistance depends on the relative orientation of $\hat{m}$ and $\hat{m}_{pin}$, and undergoes oscillation if $\hat{m}$ oscillates, as follows:

$$R = R_P + \frac{\Delta R}{2}(1 - \hat{m}.\hat{m}_{pin}) \Rightarrow \delta R = -\frac{\Delta R}{2}(m_{pin,x}\delta m_x + m_{pin,y}\delta m_y) ---(4)$$

The homodyne mixture of oscillating current and resistance produces dc voltage given by:

$$V_{dc} = \frac{1}{2} I_{rf} \text{Re}(\delta R) = \frac{-1}{4} I_{rf} \Delta R [m_{pin,x} \text{Re}(\delta m_x) + m_{pin,y} \text{Re}(\delta m_y)] --- (5)$$

From above equations we see that, Oersted field term is multiplied by Re($\chi_{11}$) which has a dispersion shape whereas the anomalous magnetic field term is multiplied by Re($\chi_{12}$) which shows a peak at resonance.

Let's now see how dc voltage changes when we reverse $\hat{m}$ and/or $\hat{m}_{pin}$. If we reverse $\hat{m}$ keeping $\hat{m}_{pin}$ same, (i.e. $\theta_m \rightarrow \theta_m + \pi$), the Oersted field term remains the same as involves factors of $\sin\theta_m \times \cos\theta_m$ and $\cos^2\theta_m$, whereas the anomalous field driven term inverts as it involves factors of $\sin\theta_m$ and $\cos\theta_m$. If we reverse $\hat{m}_{pin}$ keeping $\hat{m}$ same, we see from equation 5 that $V_{dc}$ gets a minus sign. However, the anomalous field term itself changes sign if we invert $\hat{m}_{pin}$. Thus the Oersted field term changes sign, whereas the anomalous field term remain the same in this case. Combining above two scenarios, if we reverse both $\hat{m}$ and $\hat{m}_{pin}$, both the terms change sign i.e. dc voltage inverts. These conclusions are in agreement with the experimental data shown in fig 2 a-d.

If the free and pinned layer are parallel (i.e. $m_{pin,X} = \cos\theta_m$ and $m_{pin,Y} = \sin\theta_m$), the dc voltage is 0 as can be seen from equations 2,3 and 5.

We now take a particular case where the pinned layer is along y-axis. From equations 3 and 5, the dc voltage is given by:

$$V_{dc} = \frac{-1}{4} I_{rf} \Delta R \, \text{Re}(\delta m_y) = \frac{-1}{4} I_{rf} \Delta R [\cos^2\theta_m \, \text{Re}(\chi_{11}) h_{Oe} - \cos\theta_m \, \beta \, \text{Re}(\chi_{12}) J)] --- (6)$$

The above equation shows that the anti-symmetric Lorentzian (dispersion) term has $\cos^2\theta_m$ dependence whereas the symmetric Lorentzian term has $\cos\theta_m$ dependence in agreement with the experimental data in figure 4.c-d.

**S.2 Estimation of magnetization angle between free and fixed layer and its impact on ST-FMR data**

In ST-FMR experiment we have swept frequency for a particular dc magnetic field which is in the range of 400 Oe to 500 Oe. The coercivity of free layer is around 40 Oe. Hence free layer almost aligns to external magnetic field. For example at 450 Oe, the estimated maximum difference in between free layer and external field directions is less than 2.5 degree. The magnetization direction of the fixed layer can change a bit when external magnetic field is applied which can be estimated from the pinning strength. The magnetization of the stack measured by Kerr rotation (fig S2.a) indicates that the pinning field strength is about 1.6 kOe. The angular dependence of the symmetric and anti-symmetric components of the dc voltage shown in fig 4 (and fig S2 b,c) would show some deviation from $\cos\theta_m$ and $\cos^2\theta_m$ dependence if fixed layer moves. We have numerically evaluated the angular dependence taking account the rotation of fixed layer which is shown by the blue curve in fig S2. b-c. (The red curve shows $\cos\theta_m$ and $\cos^2\theta_m$ dependence.) The experimental results (black data points) are well described by the blue curve.

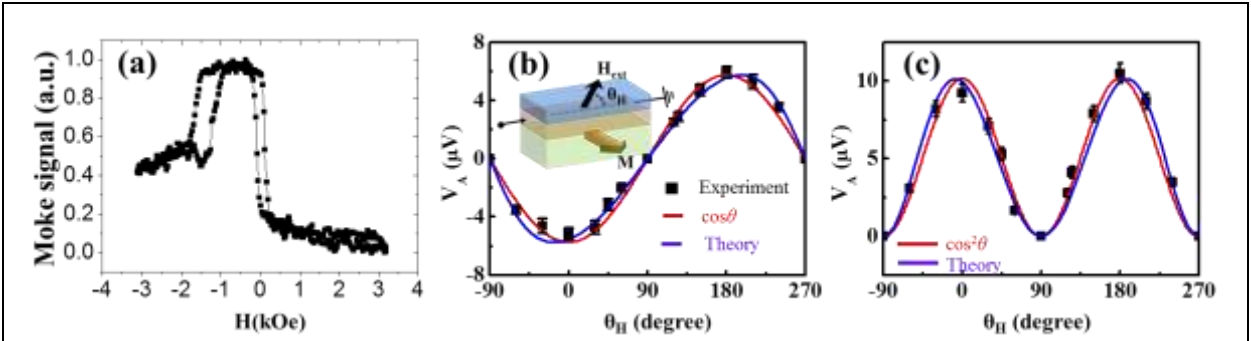

**Figure S2**| (**a**) Magneto-optic kerr measurement of the GMR stack (**b**), (**c**) Angular dependence of symmetric ($V_S$) and anti-symmetric ($V_A$) component of ST-FMR experiment respectively. Red curve in figure (**b**) and (**c**) indicates $\cos\theta$ and $\cos^2\theta$ dependence respectively. Blue curve in figure (**b**) and (**c**) is obtained numerically taking into account rotation of fixed layer due to finite exchange bias (~1.6 kOe).

### S.3 Numerical evaluation of the dc voltage signal:

We can numerically evaluate the dc voltage from equation in section S1. The Oersted magnetic field depends on the current distribution in the sample, which can be obtained from the electrical conductivities of the various layers in the stack. The anomalous magnetic field is given by, $\bar{h} = \beta(\hat{m}_{pin} \times \bar{J})$ where J is taken as average current density flowing in the free layer/Cu spacer/pinned layer stack and $\beta$ is taken as a parameter to be evaluated.

Following parameters are used for numerical calculation.

| Metal | Thickness (nm) | Resistivity (ohm-m) |
| --- | --- | --- |
| Cu (cap) | 3 | 8E-8 |
| CoFe (free layer) | 2 | 2.7E-7 |
| Cu (spacer) | 5 | 8E-8 |
| CoFeB (pinned) | 2 | 2.7E-7 |
| Buffer layer (IrMn(7)+R(5)+Ta(5)) | 17 | 2.5E-7 (equivalent) |

Length ($L_x$) of GMR sample is 375 μm and width is 25 μm. Applied field is 450 Oe. Our GMR sample shows high out of plane anisotropic field (~1.35E4 Oe) and higher damping ($\alpha$=0.09). Chosen resistivity closely matches the experimental and simulated resistance of GMR stack and magneto resistance. Experimentally obtained resistance of in-plane GMR stack is around 88 ohm whereas simulated result of GMR resistance is 85.5 ohm. The rotation of the fixed layer on application of magnetic field is also taken into account in the numerical calculation. It is found that $\beta \approx -10^{-8}$ m gives a reasonable match to the experimental data as shown in Figure S3.a-b for two different configurations. (Same experimental data as in fig 2 a,b) This value can be written as $\beta \approx 125$ Oe/$10^{12}$ (A/m$^2$) i.e. an average current density of $10^{12}$ A/m$^2$ passing through

CoFe(2nm)/Cu(5nm)/CoFe(2nm) layers, produces a magnetic field of 125 Oe along Z-direction on the free layer if the pinned layer and current are perpendicular to each other.

Figure S3.c shows the configuration when fixed layer is along *X*-axis and free layer is along *Y*-axis (same as fig 4c in the main paper). The expected dc voltage is zero if we assume that the pinned layer does not rotate i.e. remains along *X*-axis, when magnetic field along *Y*-axis is applied. Experimentally, we do observe a small dc voltage (compare Fig S3.a and Fig S3.c) in this configuration which can be explained by the rotation of fixed layer as argued in S.2 section. Numerical calculation (red curve in Fig S3.c) including 1.6 kOe exchange bias can reproduce the experimental data of figure S3.c. Figure S3.d shows the experimental data when both free and pinned layer are along x-axis (parallel to current) for two different values of external magnetic field (300Oe and 500 Oe). In both these cases we see no signal from ST-FMR since our detection method is based on in-plane GMR. However at low frequency range (around 1 to 2 GHz) small peak (amplitude less than 1.5 µV) appears (in all cases) which are magnetic field independent (compare figure S3.a,b and Fig S3.d). This background voltage along with a constant dc background is subtracted from experimental data shown in main article. Typically resonance frequencies in our experiment are the range of 5-6 GHz which allows us to detect ST-FMR signal with minimal error.

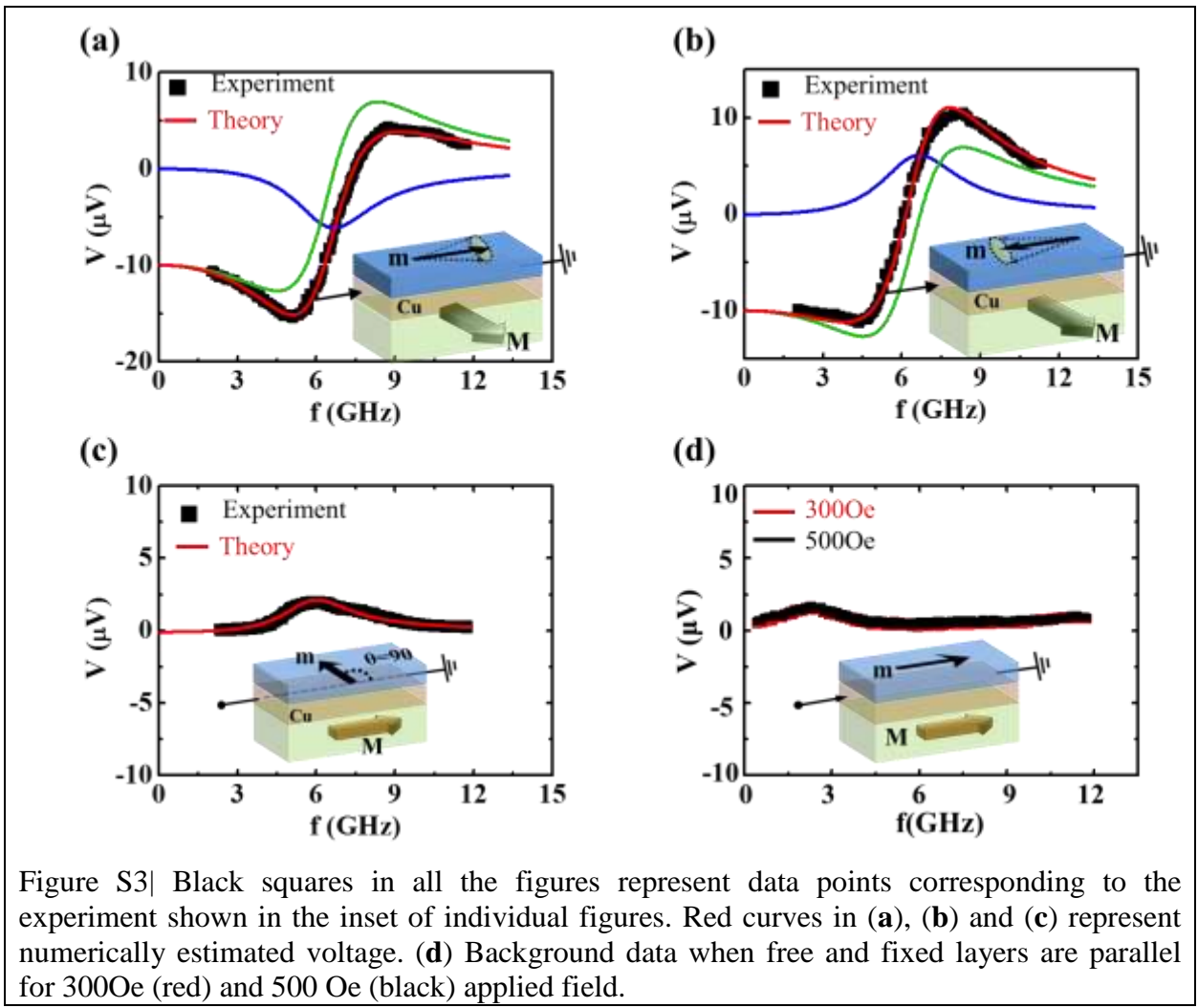

Figure S3| Black squares in all the figures represent data points corresponding to the experiment shown in the inset of individual figures. Red curves in (**a**), (**b**) and (**c**) represent numerically estimated voltage. (**d**) Background data when free and fixed layers are parallel for 300Oe (red) and 500 Oe (black) applied field.